# The Future of Virtual Classroom: Using Existing Features to Move Beyond Traditional Classroom Limitations

Michalis Xenos

University of Patras, Computer Engineering and Informatics Department
`xenos@upatras.gr`

**Abstract.** This paper argues that the true potential of virtual classrooms in education is not fully exploited yet. The features available in most environments that have been incorporated as virtual classrooms are classified into two groups. The first group includes common features, related only to the emulation of a traditional classroom. In this group, the practical differences between traditional and virtual classroom are discussed. In addition, best practices that could aid the professors to make students feel like participating in a typical classroom are presented. The second group comprises of advanced features and practices, which extend the traditional classroom. In this group, examples of successful practices which could not be performed in a traditional classroom are introduced. Finally, a qualitative study with interviews of 21 experts from 15 countries is presented, showing that even these experts are not fully exploiting the advanced features that contemporary virtual classroom environments are offering.

**Keywords:** eLearning, Virtual Classrooms

## 1   Introduction

Virtual classrooms today are used by educators to replicate a customary practice carried out for centuries, i.e. to teach exactly as they did in a typical classroom. In most cases, this is exactly what learners anticipate, leading virtual classroom usage into a vicious cycle. Although the technology is available for virtual classrooms to move beyond traditional (face-to-face) educational settings and to include practices that cannot be carried out in a traditional classroom, this is not the case and it will probably take some more years to become a widespread practice. This situation is similar in many cases when a modern technology is introduced in a field with established methods and traditions for numerous years.

For example, similar approach has been adopted film industry. While the history of movies began in 1890, all pioneer movies emulated what people (audience and actors) knew from centuries ago: theater. Therefore, the first movies were filmed with stage-bound cameras, the actors did what they knew best before movies, acting on the stage, while the scenes were assumed to follow a linear chronological succession. The first movie to truly explore some of the potential, which the new medium could offer was the "*Great Train Robbery*" filmed in 1903, where for the first-time location shooting and events happening continuously at identical times but in different places were introduced to the audience [1]. It required a lot of time for all the films to adopt to such practices, which nowadays are common in film making.

This paper argues that while virtual classrooms could move beyond traditional classroom limitations, their usage is still bounded by 'tradition', as in the film industry paradigm. To present this case, the following section includes a brief literature review of virtual classrooms. Section 3 presents experiences and best practices from using virtual classrooms either just to emulate traditional ones, or attempting to move beyond traditional classroom limitations. Section 4 presents a qualitative study with interviews of 21 educators, experts in using virtual classrooms for higher education, which illustrates the ascertainment that even these educators do not fully exploit these features. Finally, section 5 summarizes the main findings of this paper.

## 2  Virtual classrooms for eLearning

The use of networked computers to enhance learning was introduced as early as 1980, when Chambers [2] suggested that distance learning experiments should be implemented in a way that could enable in-house learning for some educational materials. The term virtual classroom was introduced in 1986, when Hiltz [3] perceived the use of a computerized conference system as a "*virtual classroom*". The early uses of virtual classrooms focused on practical issues such as sound and video and use of a "*pencil*" for the whiteboard, while the main problems reported were related to limited bandwidth and lack of "*turn-taking*" [4, 5]. As soon as video conference technology evolved and matured, a lot of online synchronous tools for learning have been emerged offering choices for virtual classrooms [6]. Most environments offered features like real-time voice and video, whiteboard, slides presentation, text-based interaction and means for learners' feedback [7].

The use of virtual classrooms was initially driven by necessity, mainly in the context of synchronous distance learning, where a professor had to emulate a typical classroom for distance students. In these early examples, the main goal was to succeed to offer students an experience similar to a face-to-face classroom. In many cases this wasn't succeeded, due to network and equipment limitations that lead into sound and video problems, as well as due to lack of suitable tools (i.e. a discussion administration feature). As new environments started to include more features [6], leaving the sound and video issues in the past, focus was given into the quality and the usability of the environment [8-12]. Using virtual classrooms wasn't only something for distance learners, but also for blended learning, or even as a supplement of on-campus courses [13].

Nowadays, within a virtual classroom, synchronous communication between distance learners may be used to better support personal participation, inducing arousal and motivation [14] and help students to better form a learning community and avoid alienation, which is inversely related to classroom community [15]. Assignments involving collaboration in virtual classroom groups increase the efficiency of the learning process as well as student competencies [16]. In contemporary virtual classroom environments, there is a variety of features available that could be exploited not only to emulate a traditional classroom, but also to move beyond the traditional classroom limitations.

## 3  Experiences from using virtual classrooms

To present better the author's experiences from twenty years of using virtual classrooms in the tertiary education, the features available in most virtual classroom environments are classified into two groups. The first group (common features) includes features related only to the emulation of a traditional classroom. The second group (advanced features) comprising of features and practices going beyond the traditional classroom. Table 1, includes both categories.

Table 1. Common and advanced features and practices in virtual classrooms

| Common Features | Advanced Features |
|---|---|
| Video and sound | Retrospective assignments |
| Chat | Breakout rooms |
| Students' feedback | Anonymous polling |
| Whiteboard | Shared whiteboard |
| Slide presentations | Shared documents and annotating |
| Discussion administration | Application sharing |

To distinguish the advanced features, the following requirements had to be met: a) being available in most virtual classroom environments that have been incorporated up today, b) being documented extensively, so most educators could be familiarized with these features, c) being available only online and not in a traditional classroom, at least not without having to overcome physical and practical limitations. In the following subsections, the practical differences between traditional and virtual classroom are discussed and experiences from the use of these common features are presented, while examples of best practices related to advanced features that could not be performed in a typical traditional classroom are introduced.

### 3.1  Using common features of a virtual classroom to emulate a traditional one

Nowadays **video and sound** is available for both the professor and the students (or at least sound from all the students). However, this was not the case for the virtual classrooms at the beginning of this century. Video from students increases the sense of community and the best practice is to try to have all students present themselves on video, especially in cases that they haven't met face-to-face. It is a fact that in distance education, having students met at least once is valuable for building a community [17] and in cases that this was not feasible, allowing them to introduce themselves using video and sound is essential.

A **chat feature** can always help overcoming sound problems and, although is not related to a traditional classroom practice, is also included in this list for both histori-cal and practical reasons. Although this is not something occurring very often today, reviewing recordings from 2000 to 2003 reveals that almost one third of the students participating in virtual classrooms faced sound problems during a session [18]. The chat feature, apart from solving sound issues could allow students to better clarify a question, or to allow the professor to collect short responses, especially if the chat supports direct student-to-professor messages, as most contemporary environments do. Experience had revealed that, although action in the chat is a measure of active participation, a single professor is unable to handle both oral and written communication. In this case, a solution is having two educators present (one responsible from collecting chat messages and presenting them orally). When the chat system is expected to be used by students and the number of participants is higher than 30, having more than one professor present is strongly advised.

The **students' feedback feature** allows the professor to effectively monitor participation. Expecting from each student to take a turn and reply to a simple question "is everything OK so far?", or "can everyone hear me?" may take several minutes and distress the normal flow of the lecture, while goes naturally using the feedback mechanism. Most environments offer complex feedback, including emotions, but the best advice is to keep it simple to a "yes" and "no". Normally, the best practice is asking for a confirmation at least every 10 to 15 minutes, usually following this confirmation with an activity that will further involve students.

The **whiteboard feature**, while in a typical classroom is always a problem for the professor, since extensive use requires to turn the back to audience for a long time, is a major asset in virtual classrooms. The more a professor uses the whiteboard, the more engages the students and the best practice is to frequently allow students to write on whiteboard, or to let them highlight areas they want to discuss further.

The **slide presentations feature** is a valuable feature, only if used with caution. Ideally one should only use slides with complex schemas and images. A virtual classroom based entirely on slide presentation turns out to be a webcast. The best practice is that if something can be sketched in the whiteboard, use the whiteboard instead of a slide. When using a slide is unavoidable, use the virtual laser pointer, add comments, ask students to point, or highlight and do anything possible to engage students into the discussion.

Finally, the **discussion administration feature** facilitates the most challenging task the professor using a virtual classroom must tangle. Controlling the audience, monitoring the 'raised hands' and allowing 'turns' to speak, is something one need to practice for a while, before mastering it in practice. Since it is a quite often phenomenon that some students will 'raise hand' and then cancel it, especially the shy ones, it takes practice from the professor to be able to control the flow of the discussion and do not let students feel left out. The best practice is when the audience is under 10 students to set all microphones on and disable the 'hand raise' feature, while for larger audience using it is required. In some extreme cases of many participants, having an assistant to monitor raised hands from students can be proven extremely helpful (especially when the audience is above 50 to 60 students). In case that after speaking for 10 to 15 minutes there are no hands raised, the best practice is to take a break and ask something to engage the audience.

### 3.2 Advanced features for going beyond the traditional classroom

Since all the virtual classroom sessions can be recorded and viewed many times, the recording feature can be used for **retrospective assignments**. The best practice, exploiting the recording feature and forcing students to review a session, is to relate assignments to the previous session. Assignments like "*In the 25th minute of the session, a student asked about …. After hearing the discussion, could you offer some more options?*", require from students to review the recorded session and is a valuable educational practice.

The feature of **breakout rooms** is very powerful for engaging students. There are a variety of teamworking practices that could exploit this feature and using it properly could really enhance the educational experience. While in typical classrooms such teamworking is always in terms with physical limitations, in virtual classrooms is something that can be done with ease. The best practice is to engage students using breakout rooms quite frequently during a virtual classroom session (at least once in every session) and to have students report back to the main room their discussion.

Although the technology for **anonymous polling** could also exist in a typical classroom, this requires equipment not commonly available, while it is common in all virtual classroom environments. The use of anonymous polling, engages students and provides the professor with real time feedback. Having one or two review questions every 10 minutes is the best practice. While the typical student feedback may be 100% "*yes*" in the question "*is everything clear so far*", a couple of review questions may reveal the need to repeat a part of the session, or to start a discussion. In a typical classroom, usually one student will reply correctly to these questions and the rest will silently concur, misleading the professor to think that everything was understood.

The feature of the **shared whiteboard** allows students' participation in activities related to design charts, graphs and similar. Having a few students working together on the whiteboard for a task is a valuable practice. Usually, in the main room the professor could ask for 2 and up to 4 volunteers to work on an exercise, but the best practice is to use a breakout room, allowing a small number of students in each room (depending on the activity 2 and up to 4, or even 5 students). Some virtual classroom environments allow the results of each room whiteboard to be shared back to the main room, but not all. If this feature is not available, usually a working solution is to use a print-screen of the results to report it back to the main room.

The feature of **shared documents and annotating**, allows the professor to have students working together on a document, i.e. reviewing code and annotating as part of a collaborative exercise. Depending on the number of students, this is something that can be done in the main room (usually when up to 10 students attend the virtual classroom), or using breakout rooms. The best practice for larger audiences is to combine breakout rooms with such collaborative exercises.

The feature of **application sharing** is important, not only for the professor sharing an application to demonstrate the use of a software tool, but also for students. In fact, having students share their application to present a problem while the professor comments on that is a powerful and constructive educational experience. In the field of computer science, where students are required to use many software tools, this prac-tice speeds up significantly the process of responding to questions and providing appropriate feedback. It is much easier to view the students' solution and comment on it, rather than having them explain their solution and making assumptions. This is also a very helpful educational practice for all the students participating in the virtual classroom, as long as the discussion is not monopolized on a single student's solution. This practice could be also valuable when a virtual classroom session is used as 'office hours' for responding to students' questions.

## 4    A survey on the exploitation of the advanced features

For the six advanced features that could aid the virtual classrooms to go beyond traditional classroom limitations and fully exploit what technology offers, an informal interview was conducted involving 21 educators from the following 15 countries: Austria, Cyprus, Czech Republic, Belgium, Germany, Greece, Finland, Italy, Lithuania, Poland, Portugal, Spain, The Netherlands, Turkey, and the United Kingdom. All the interviewed educators are experts in using virtual classrooms for higher education. The interviews were informal, feeling more like a friendly discussion, trying to minimize note taking and allowing the discussion to include successful experiences from the virtual classroom usage, or anecdotes of failures. The questions asked were the following, starting with "Have you…":

- Q1: … assigned something that would require from students to review the session from the lectures archive?
- Q2: … used breakout rooms to let the students work on a collaborative assignment?
- Q3: … used anonymous online polls during the lecture?
- Q4: … shared the whiteboard to more than one student at the same time?
- Q5: … used a shared document and asked students to annotate?
- Q6: … asked from students to share an application to demonstrate a problem?

The frequency ranges were informally discussed and sometimes the interviewees failed to provide a clear answer or gave answers like "*I don't know if it is five or ten, maybe less than five, maybe closer to ten, but definitely isn't something that fits in my classroom*", so the frequencies are presented as follows:

- F0: Never, or just to test the tool but not in a real classroom.
- F1: A few (one to five) times over all the years, but this never became a customary practice.
- F2: Not frequently, but sometimes and not on every course I teach.
- F3: Frequently (more than once in each course I teach), but this is not a regular process.
- F4: This is a regular process I use in my virtual classroom sessions.

The results for each of the six features are presented in Table 2, where the six rows correspond to the features and the five columns to the usage frequency. Even though most (12 out of 21) educators are from the technology field (teaching STEM courses) and therefore are expected to be familiar with the use of modern virtual classroom features, results showed that in most cases the features were tested and never actually exploited in practice. In fact, considering that the results could be biased towards a positive attitude of the technology –since most of the educators were tech savvy, which made more difficult for them to admit that they didn't use these features– the results indicate that exploitation of the advanced features of virtual classrooms hasn't reached its full potential yet. Some features, like the retrospective assignments start to become part of normal practice, others like the shared whiteboard and documents are included occasionally into some sessions, while other like the anonymous polls, the application sharing and the breakout rooms are just starting to be acknowledged as promising opportunities.

Table 2. Results from the survey

|    | F0 | F1 | F2 | F3 | F4 |
|----|----|----|----|----|----|
| Q1 | 6  | 4  | 7  | 2  | 2  |
| Q2 | 20 |    | 1  |    |    |
| Q3 | 15 | 4  | 1  |    | 1  |
| Q4 |    | 6  | 11 | 3  | 1  |
| Q5 | 7  | 9  | 5  |    |    |
| Q6 | 19 | 2  |    |    |    |

## 5 Conclusions

This study suggests that the use of virtual classrooms hasn't reached its full potential yet and there are features that could be employed to aid towards moving virtual classrooms beyond just emulating traditional ones. The paper presents examples of best practices using such advanced features, based on the author's experience. It also presents practices that could be used to improve teaching, based on the common features of virtual classrooms, which are used mostly to emulate teaching as in traditional classrooms. Nowadays most professors still use virtual classrooms to replicate the practice they are familiar with: teaching in a face-to-face classroom. This is typical when technology is introduced in a practice which exists for many years. The best practices presented in this paper could aid professors to move beyond traditional classroom limitations and fully exploit the entire spectrum of modern virtual classroom features.

## References

[1] T. Dirks. "Filmsite Movie Review, The Great Train Robbery (1903)," June 2, 2017; http://www.filmsite.org/grea2.html.
[2] J. A. Chambers, and J. W. Sprecher, "Computer assisted instruction: current trends and critical issues," *Commun. ACM,* vol. 23, no. 6, pp. 332-342, 1980.
[3] S. R. Hiltz, "The "virtual classroom": Using computer-mediated communication for university teaching," *Journal of communication,* vol. 36, no. 2, pp. 95-104, 1986.

[4]  D. Dwyer, K. Barbieri, and H. M. Doerr, "Creating a virtual classroom for interactive education on the Web," *Computer Networks and ISDN Systems,* vol. 27, no. 6, pp. 897-904, 1995.

[5]  S. R. Hiltz, and B. Wellman, "Asynchronous learning networks as a virtual classroom," *Commun. ACM,* vol. 40, no. 9, pp. 44-49, 1997.

[6]  S. Schullo, A. Hilbelink, M. Venable, and A. E. Barron, "Selecting a virtual classroom system: Elluminate live vs. Macromedia breeze (adobe acrobat connect professional)," *MERLOT Journal of Online Learning and Teaching,* vol. 3, no. 4, pp. 331-345, 2007.

[7]  J. Finkelstein, "Learning in real time," *San Francisco: Jossy-Bass Publishing Company. CA*, pp. 94103-1741, 2006.

[8]  M. Xenos, and D. Christodoulakis, "Software Quality: The user's point of view," *Software Quality and Productivity: Theory, practice, education and training*, M. Lee, B.-Z. Barta and P. Juliff, eds., pp. 266-272, Boston, MA: Springer US, 1995.

[9]  J. Johnston, J. Killion, and J. Oomen, "Student satisfaction in the virtual classroom," *Internet Journal of Allied Health Sciences and Practice,* vol. 3, no. 2, pp. 6, 2005.

[10]  D. Stavrinoudis, M. Xenos, P. Peppas, and D. Christodoulakis, "Early Estimation of Users' Perception of Software Quality," *Software Quality Journal,* vol. 13, no. 2, pp. 155-175, 2005.

[11]  A. Stefani, B. Vassiliadis, and M. Xenos, "On the quality assessment of advanced e-learning services," *Interactive Technology and Smart Education,* vol. 3, no. 3, pp. 237-250, 2006.

[12]  C. Katsanos, N. Tselios, and M. Xenos, "Perceived Usability Evaluation of Learning Management Systems: A First Step towards Standardization of the System Usability Scale in Greek," in 16th Panhellenic Conference on Informatics, PCI2012, 2012, pp. 302-307.

[13]  H. Singh, "Building effective blended learning programs," *Educational Technology-Saddle Brook Then Englewood Cliffs NJ-,* vol. 43, no. 6, pp. 51-54, 2003.

[14]  S. Hrastinski, "The potential of synchronous communication to enhance participation in online discussions: A case study of two e-learning courses," *Information & Management,* vol. 45, no. 7, pp. 499-506, 2008.

[15]  A. P. Rovai, and M. J. Wighting, "Feelings of alienation and community among higher education students in a virtual classroom," *The Internet and Higher Education,* vol. 8, no. 2, pp. 97-110, 2005.

[16]  A. Crişan, and R. Enache, "Virtual Classrooms in Collaborative Projects and the Effectiveness of the Learning Process," *Procedia - Social and Behavioral Sciences,* vol. 76, pp. 226-232, 2013.

[17]  A. Marsap, and M. Narin, "The integration of distance learning via internet and face to face learning: Why face to face learning is required in distance learning via internet?," *Procedia - Social and Behavioral Sciences,* vol. 1, no. 1, pp. 2871-2878, 2009.

[18]  M. Xenos, and A. Skodras, "Evolving from a Traditional Distance Learning Model to e-Learning," in 2nd International LeGE-WG Workshop on e-Learning and Grid Technologies: A Fundamental Challenge for Europe, Paris, France. 3rd & 4th March 2003, Paris, France, 2003, pp. 121-125.